\documentclass[a4paper,11pt]{article}
\pdfoutput=1 % if your are submitting a pdflatex (i.e. if you have
\usepackage{jheppub} % for details on the use of the package, please
\usepackage{amsthm}

\usepackage[T1]{fontenc} % if needed
\usepackage{ifthen}
\newcommand{\Li}{\mathrm{Li}}
\newcommand{\e}{\epsilon}
\newcommand{\bs}[1]{\boldsymbol{#1}}

\newcounter{example}

\newcommand{\example}[1]{
	\stepcounter{example}
	\subsection*{
		Example \arabic{example}\ifthenelse{\equal{#1}{}}{.}{: #1.}
	}
}

\title{Polylogarithmic functions with prescribed branching locus and linear relations between them.}

\author{Roman N. Lee}

\affiliation{Budker Institute of Nuclear Physics, Novosibirsk 630090, Russia}

\emailAdd{r.n.lee@inp.nsk.su}

\abstract{We consider the problem of finding the set of classical polylogarithmic functions $\Li_n$ with branching locus determined by the solution of $p_1\cdot p_2\cdot \ldots \cdot p_n=0$, where $p_1,\ldots, p_n$ are irreducible polynomials of several variables. We present an algorithm  of constructing a complete set of possible arguments of $\Li_n$ functions. The corresponding \textit{Mathematica} code is included as ancillary file. Using this algorithm and the symbol map, we provide some examples of polylogarithmic identities.}

\begin{document}
\maketitle
\flushbottom
\section{Introduction}

The problem of simplification of expressions involving classical and generalized polylogarithms often arises in the area of multiloop calculations. In particular, using IBP reduction and reduction to $\e$-form it is often possible to reduce the problem of multiloop calculations to the solution of differential system \cite{Henn2013,Lee:2014ioa}
\begin{equation}\label{eq:eform}
	\frac{\partial}{\partial x_i}\bs{J} = \e S_i(\bs{x}) \bs{J}\,,
\end{equation}
where $S_i(\bs{x})$ are the matrices with entries being the rational functions of $\bs{x}$ and $\e$ is the parameter of dimensional regularization.
Then the singular locus of the system (which corresponds to the branching locus of its solution) is defined by the equation
\begin{equation}\label{eq:singular_locus}
	\prod_{k=1}^{n}p_k(\bs{x})=0,
\end{equation}
where $p_k$ are irreducible denominators of $S_i$.

The perturbative in $\e$ solution of the system \eqref{eq:eform} is expressed in terms of Chen's iterated path integrals \cite{chen1977iterated} which in many cases can be rewritten via classical polylogarithms. The solution often has a rather cumbersome form, and the question of its simplification naturally arises. This task requires the use of various functional identities between polylogarithms. Symbol map \cite{goncharov2010classical,Duhr:2011zq} gives a natural tool for checking such identities, and also for the search of those identities provided an appropriate set of functions is known.

In the present paper we describe an approach for finding such a set of functions. Our approach provides an algorithm for finding all arguments of polylogarithmic functions $\Li_n$ with the branching locus defined by polynomial equations. We demonstrate the efficiency of our method on several examples.

\section{Polylogarithmic functions with prescribed branching locus}
Let us formulate the problem as follows.
Denote by $V$ the singular locus --- the set of solutions of Eq. \eqref{eq:singular_locus} where $p_k$ are some irreducible polynomials.

Our goal is to construct all possible rational functions
\begin{equation}
	Q(\bs{x})=N(\bs{x}) /D(\bs{x})\,,\qquad \text{GCD}(N(\bs{x}),D(\bs{x}))=1\,,
\end{equation}
such that the branching locus of the function $\Li_n(Q)$ is a subset of $V$. Recalling that the branching points of $\Li_{n>1}(z)$ are $0$, $1$, and $\infty$, we reformulate our requirement as that the solution of \textbf{each of the three equations}
\begin{equation}
	Q=0,\qquad Q=1,\qquad Q=\infty\,
\end{equation}
	is a subset of $V$. These equations can be rewritten as
\begin{equation}
	N=0,\qquad N-D=0,\qquad D=0\,.
\end{equation}
	The requirement is then equivalent to
\begin{equation}
	N= c_1 \prod_{k=1}^{n}p_k^{m_k}\quad\And\quad
	N-D= c_2 \prod_{k=1}^{n}p_k^{m_k}\quad\And\quad
	D= c_3 \prod_{k=1}^{n}p_k^{d_k},
\end{equation}
where $c_1,c_2,c_3$ are some constants and $n_k,m_k,d_k\in \mathbb{Z}_+$.

Now it is clear how we can search for the possible arguments of $\Li_n$.
\begin{enumerate}
	\item First, we construct a set of polynomials which are the products of powers of $p_k$:
\begin{equation}\label{eq:products}
	P_0=1,\ P_1=p_1,\ \ldots,\ P_n=p_n,\ P_{n+1}=p_1^2,\ P_{n+2}=p_1p_2,\ \ldots
\end{equation}
We should stop at sufficiently high overall degree.
\item Then we search for triplets of linearly dependent polynomials $P_i,P_j,P_k$, so that
\begin{equation}\label{eq:triplet}
	a_1 P_i + a_2 P_j+a_3P_k=0\,,
\end{equation}
where $a_1,\, a_2,\, a_3$ are some constant coefficients.
\item For each triplet we have 6 possible arguments of polylogarithm:

\begin{equation}
	z=-\frac{a_1 P_i}{a_2 P_j},\qquad
	-\frac{a_2 P_j}{a_1 P_i},\qquad
	-\frac{a_3 P_k}{a_2 P_j},\qquad
	-\frac{a_2 P_j}{a_3 P_k},\qquad
	-\frac{a_3 P_k}{a_1 P_i},\qquad
	-\frac{a_1 P_i}{a_3 P_k}.
\end{equation}
These arguments are related by the Moebius transformations which permute the points $0,1,\infty$, namely, by
\begin{equation}\label{eq:S3}
	\mathbb{S}_3=\left\{z\to z,\ z\to \frac{1}{z},\ z\to 1-z,
	\ z\to \frac{1}{1-z},\ z\to 1-\frac{1}{z},
	\ z\to \frac{z}{z-1}\right\}.
\end{equation}
\end{enumerate}

One might wonder if the number of valid arguments $Q=N/D$ is finite, and if it is, is there an upper bound for the number of polynomials $P_k$. The answer to both questions is positive. This follows from the extension of Stothers-Mason theorem \cite{10.1093/qmath/32.3.349,mason2006equations}, which is also known for being a precursor of the celebrated ABC hypothesis. In particular, theorem 1.2 of Ref. \cite{shapiro1994extension} restricted to the case of interest claims that the degree of the three polynomials $P_i,\ P_j,\ P_k$ selected from the set \eqref{eq:products} and satisfying \eqref{eq:triplet} is restricted by
\begin{equation}
	\deg P_{i,j,k} < \sum_{k=1}^{n}\deg p_k,
\end{equation}
Therefore, in order to find all valid arguments, we should examine a finite set of triplets.

In the ancillary file \textit{Arguments.wl} we provide an implementation of the described approach as the \textbf{\textit{Mathematica}} function \texttt{PolyLogArguments[\{$p_1,\ldots,p_n$\},\{$x_1,\ldots,x_m$\}]} which finds all possible arguments of polylogarithmic functions with branching locus defined by Eq. \eqref{eq:singular_locus}. The result of this function is a list of sextets of arguments with each sextet being the orbit of the group defined in Eq. \eqref{eq:S3}. The examples of using this function are provided in the ancillary file \textit{Examples.nb}.

%\section{Using symbol map to discover identities}
%As we explained in the previous section, it is possible to find a complete set of polylogarithmic functions with a prescribed branching locus.
%What identities can we expect among the polylogarithmic functions? It can be proved that the identities with constant coefficients constitute a basis of all identities. That is, suppose that we have an identity of the form
%\begin{equation}
%	\sum_{k=0}^{N} R_k(\bs{x}) F_k(\bs{x})=0,
%\end{equation}
%where $R_k(\bs{x})$ are \textbf{linearly independent} rational functions and $F_k(\bs{x})$ are the linear combinations of (products) of polylogarithms with constant coefficients. Then necessarily
%\begin{equation}
%F_k(\bs{x})=0 \text{ for all $k$.}
%\end{equation}
%
%A ``physical'' explanation of this fact is that polylogarithms, similar to logarithms, are ``slowly varying functions'' in contrast to the rational coefficients $R_k$, so the identity may hold only if each separate term is zero. A rigorous proof of this statement for the case of one variable is presented in the appendix.
%
%In order to prove the statement in a rigorous way, we use induction over the maximal transcendental weight and the number of terms with maximal transcendental weight. The base of the induction for zero transcendental weight is obvious: $\sum_{k=0}^{N} R_k(\bs{x}) c_k=0$ for linearly independent $R_k$ holds iff $c_k=0$.
%
%Suppose now that the theorem holds for all expressions with at most $m$ terms of maximal transcendental weight $n$.

\section{Examples}

Let us consider some examples of applying the above approach.
In some of the examples below we will use Lewin's notation, \cite[Eq. (3.19)]{lewin1991structural},
\begin{equation}\label{eq:L_n}
	L_n(z)=\Li_n(z)+\sum_{r=1}^{n-2} \frac{(-1)^r}{r!} \ln^r |z|\, \Li_{n-r}(z)+(-1)^n\frac{n-1}{n!}\ln^{n-1}|z|\,\ln (1-z)
\end{equation}
for real $z$ less than one. Below we will assume also that all variables are real and vary from $0$ to $1$ unless otherwise stated.

\example{trivial case}
Let us take
\begin{equation}
	p_1=x,\ p_2=1-x.
\end{equation}

Our algorithm delivers an expected result for all possible arguments:
$$\left\{x,\frac{1}{x},1-x,\frac{1}{1-x},\frac{x-1}{x},\frac{x}{x-1}\right\},$$
corresponding to the action of $\mathbb{S}_3$ group \eqref{eq:S3}.
In the following examples not to clutter the presentation we will present a list of arguments modulo the action of this group.

\example{additional branching point $x=-1$}

Let us now take
\begin{equation}
	p_1=x,\ p_2=1-x,\ p_3=1+x
\end{equation}
Our algorithm, up to the action of $\mathbb{S}_3$ group in Eq. \eqref{eq:S3}, gives 6 possible arguments:
\begin{equation}
	\left\{x,-x,x^2,\frac{1-x}{1+x},-\frac{1-x}{1+x},\left(\frac{1-x}{1+x}\right)^2\right\}
\end{equation}
For weight 2 we then have the following list of functions:
\begin{gather}
	\bigg\{\Li_2(x),\Li_2(-x),\Li_2\left(x^2\right),\Li_2\left(\frac{1-x}{1+x}\right),\Li_2\left(-\frac{1-x}{1+x}\right),\Li_2\left(\frac{(1-x)^2}{(1+x)^2}\right),\\
	\ln ^2(x),\ln (1-x) \ln (x),\ln (x) \ln (1+x),\ln ^2(1-x),\ln (1-x) \ln (1+x),\ln ^2(1+x)\bigg\}
\end{gather}

Using symbol map we obtain two elementary identities of the same form
\begin{equation}
\Li_2\left(z^2\right)-2 \Li_2(-z)-2 \Li_2(z)=0
\end{equation}
with $z=x$ and $z=\frac{1-x}{1+x}$ and one less trivial identity
\begin{equation}
%id2.2
4 \Li_2\left(\tfrac{1-x}{2}\right)+4 \Li_2\left(-\tfrac{1-x}{2 x}\right)-2 \Li_2\left(-\tfrac{(1-x)^2}{4 x}\right)+\ln ^2(x)
=0
%id2.2/
.
\end{equation}

For weight 3 we find, for example, an identity
\begin{multline}\label{eq:idL3}
%id2.3
	L_3\left(\tfrac{1-x}{2}\right)-\tfrac{1}{4} L_3\left(\tfrac{-4 x}{(1-x)^2}\right)+L_3\left(\tfrac{-2 x}{1-x}\right)-\tfrac{1}{4} L_3\left(\tfrac{4 x}{(1+x)^2}\right)+L_3\left(\tfrac{2 x}{1+x}\right)+L_3\left(\tfrac{1+x}{2}\right)= \tfrac74\zeta_3
%id2.3/
	\,,
\end{multline}
where $L_n$ is defined in Eq. \eqref{eq:L_n}. Note that, using identities for $\Li_2$, we can eliminate all $\Li_2$ in the above identity and obtain
\begin{multline}\label{eq:idhL3}
	\hspace{-5mm}
%id2.3t
	\widetilde{L}_3\left(\tfrac{1-x}{2}\right)-\tfrac{1}{4} \widetilde{L}_3\left(\tfrac{-4 x}{(1-x)^2}\right)+\widetilde{L}_3\left(\tfrac{-2 x}{1-x}\right)-\tfrac{1}{4} \widetilde{L}_3\left(\tfrac{4 x}{(1+x)^2}\right)+\widetilde{L}_3\left(\tfrac{2 x}{1+x}\right)+\widetilde{L}_3\left(\tfrac{1+x}{2}\right)= \tfrac74\zeta_3-\tfrac32\zeta_2\ln{x}
%id2.3t/
	\,,
\end{multline}
where \begin{equation}\label{eq:tildeL}
	\widetilde{L}_3(z)=\Li_3(z)+\tfrac16\ln(1-z)\ln^2|z|-\zeta_2\ln|z|\,.
\end{equation}

For weight 4 we find 12-term relation
\begin{multline}
%id2.4
	L_4(1-x)-L_4\left(\tfrac{1+x}{2}\right)-L_4\left(\tfrac{1-x}{2}\right)-L_4\left(\tfrac{x}{x-1}\right)
	-L_4\left(\tfrac{1}{1+x}\right)-L_4\left(\tfrac{x}{1+x}\right)
	+L_4\left(\tfrac{-2 x}{1-x}\right)
	+L_4\left(\tfrac{2 x}{1+x}\right)
	\\
	-\tfrac{1}{8} L_4\left(1-\tfrac{1}{x^2}\right)
	-\tfrac{1}{8} L_4\left(1-x^2\right)
	-\tfrac{1}{8} L_4\left(-\tfrac{4 x}{(1-x)^2}\right)
	-\tfrac{1}{8} L_4\left(\tfrac{4x}{(1+x)^2}\right)
	\\
	=\tfrac{3 \zeta _2^2}{80}+\tfrac{1}{2} \zeta _2 \ln ^2{2}-\tfrac{7}{4} \zeta _3 \ln{2}-2 \text{Li}_4\left(\tfrac{1}{2}\right)-\tfrac{\ln^4{2}}{12}
%id2.4/
\,.
\end{multline}

\example{irreducible polynomial}
Let us now take
\begin{equation}
	p_1=x,\ p_2=1-x,\ p_3=1-x+x^2
\end{equation}
Our algorithm gives 4 possible arguments (mod Eq. \eqref{eq:S3}):
\begin{equation}
	\left\{x,x(1-x),-\frac{x^2}{1-x},-\frac{(1-x)^2}{x}\right\}
\end{equation}
For weight 2 we have the identity
\begin{equation}
%id3.2
	L_2\left((1-x) x\right)
	-L_2\left(-\tfrac{x^2}{1-x}\right)
	-L_2\left(-\tfrac{(1-x)^2}{x}\right)
	=\zeta_2
%id3.2/
\,.
\end{equation}
For weight 3 we have
\begin{multline}
%id3.3
	2 L_3((1-x) x)
	-L_3\left(-\tfrac{x^2}{1-x}\right)
	-L_3\left(-\tfrac{(1-x)^2}{x}\right)
	\\
	+3 L_3\left(1-x+x^2\right)
	-3 L_3\left(\tfrac{1-x}{1-x+x^2}\right)
	-3 L_3\left(\tfrac{x}{1-x+x^2}\right) = 0
%id3.3/
	\,.
\end{multline}
For weight 4 we find
\begin{multline}
%id3.4
	L_4((1-x) x)
	-L_4\left(-\tfrac{x^2}{1-x}\right)
	-L_4\left(-\tfrac{(1-x)^2}{x}\right)
	+3 L_4\left(\tfrac{1-x}{1-x+x^2}\right)
	+3 L_4\left(\tfrac{x}{1-x+x^2}\right)
	\\
	-\tfrac{3}{2} L_4\left(\tfrac{(1-x)^2}{1-x+x^2}\right)
	+3 L_4\left(-\tfrac{(1-x) x}{1-x+x^2}\right)
	-\tfrac{3}{2} L_4\left(\tfrac{x^2}{1-x+x^2}\right)
	+\tfrac{3}{2} L_4\left(1-x+x^2\right)=\tfrac{19}4\zeta_4
%id3.4/
	\,.
\end{multline}
\example{two variables}

Let us take
\begin{equation}\label{eq:blocus}
	\{p_1,\ldots, p_5\}=\{x,1-x,y,1-y,1-x y\}
\end{equation}
We obtain the following 5 arguments (mod Eq. \eqref{eq:S3})
\begin{equation}
	\left\{x,y,x y,\tfrac{x (1-y)}{1-x y},\tfrac{(1-x) y}{1-x y}\right\}
\end{equation}
Using the symbol map we obtain the celebrated 5-term identity \cite{spence1809essay}
\begin{equation}
%id4.2
	L_2(x\,y)+L_2\left(\tfrac{(1-y) x}{1-x\,y}\right)+L_2\left(\tfrac{(1-x) y}{1-x\,y}\right)-L_2(x)-L_2(y)=0
%id4.2/
	\,.
\end{equation}

We were not able to find nontrivial relations for weight 3 or higher for the branching locus defined by Eq. \eqref{eq:blocus}. However, if we add $p_6=x-y$, we find four new arguments (mod Eq. \eqref{eq:S3}):
\begin{equation}
	\left\{\tfrac{x}{y},\tfrac{y-x}{1-x},\tfrac{y-x}{(1-x) y},\tfrac{(1-y)^2 x}{(1-x)^2 y}\right\}
\end{equation}
of which the last is the most remarkable as $\Li_n\left(\tfrac{(1-y)^2 x}{(1-x)^2 y}\right)$ has branching locus on \textbf{all} surfaces $p_i=0$ with $i=1,\ldots,6$.

Then at weight 3 we discover one new 12-term identity
\begin{multline}
%id4.3
	\tfrac{1}{2} L_3\left(\tfrac{x}{y}\right)+\tfrac{1}{2} L_3(x\,y)
	-L_3(x)-L_3(y)
	+L_3\left(\tfrac{x-y}{1-y}\right)
	+L_3\left(\tfrac{y-x}{1-x}\right)
	-L_3\left(\tfrac{(1-y)x}{(1-x) y}\right)
	\\
	+\tfrac{1}{2} L_3\left(\tfrac{(1-y)^2 x}{(1-x)^2 y}\right)
	+L_3\left(\tfrac{1-x}{1-x\,y}\right)+L_3\left(\tfrac{1-y}{1-x\,y}\right)+L_3\left(\tfrac{(1-y)x }{1-x\,y}\right)+L_3\left(\tfrac{(1-x) y}{1-x\,y}\right)=2\zeta_3
%id4.3/
	\,,
\end{multline}
where $0<x<y<1$.

\example{three variables}

Finally, let us consider the set
\begin{equation}\label{eq:blocus1}
	\{p_1,\ldots, p_{10}\}=\{x,1-x,y,1-y,z,1-z,1-x\,y,1-y z,1-z x, 1-x\,y z\}.
\end{equation}
We find 22 possible arguments (mod Eq. \eqref{eq:S3})
\begin{gather}
	x,y,z, yz, zx, xy, xyz,
	-\tfrac{ (1-y)x}{1-x},
	-\tfrac{ (1-x)y}{1-x},
	-\tfrac{ (1-z)y}{1-x},
	-\tfrac{ (1-y)z}{1-x},
	-\tfrac{ (1-x)z}{1-x},
	-\tfrac{ (1-z)x}{1-x},
	\nonumber
	\\
	\tfrac{1-x}{1-x\,y z},
	\tfrac{1-y}{1-x\,y z},
	\tfrac{1-z}{1-x\,y z},
	\tfrac{(1-x) y z}{1-x\,y z},
	\tfrac{(1-y)z x }{1-x\,y z},
	\tfrac{(1-z)x\,y }{1-x\,y z},
	-\tfrac{(1-y) (1-z)x }{(1-x) (1-x\,y z)},
	-\tfrac{(1-z) (1-x)y }{(1-x) (1-x\,y z)},
	-\tfrac{(1-x) (1-z)z }{(1-x) (1-x\,y z)}.
\end{gather}
Using symbol map, we obtain the identity
\begin{multline}\label{eq:goncharov1}
%id5.3
	\tfrac{1}{6} L_3(x\,y\,z)-\tfrac{1}{2} L_3(x\,y)+\tfrac{1}{2}L_3(x)
	+\tfrac{1}{2} L_3\left(\tfrac{1-x}{1-x\,y\,z}\right)
	+\tfrac{1}{2} L_3\left(\tfrac{(1-z)x\,y }{1-x\,y\,z}\right)
	+L_3\left(-\tfrac{ (1-y)x}{1-x}\right)
	\\
	-\tfrac{1}{2} L_3\left(-\tfrac{(1-y) (1-z)x }{(1-x) (1-x\,y\,z)}\right)
%id5.3/
	+\text{permutations} =
%id5.3r
	3\zeta_3
%id5.3r/
	\,.
\end{multline}

As previously, this identity can be rewritten in the form free of $\Li_2$ functions:
\begin{multline}\label{eq:goncharov2}
%id5.3t
	\tfrac{1}{6} \widetilde{L}_3(x\,y\,z)-\tfrac{1}{2} \widetilde{L}_3(x\,y)+\tfrac{1}{2}\widetilde{L}_3(x)
	+\tfrac{1}{2} \widetilde{L}_3\left(\tfrac{1-x}{1-x\,y\,z}\right)
	+\tfrac{1}{2} \widetilde{L}_3\left(\tfrac{ (1-z)x\,y}{1-x\,y\,z}\right)
	+\widetilde{L}_3\left(-\tfrac{(1-y)x }{1-x}\right)
	\\
	-\tfrac{1}{2} \widetilde{L}_3\left(-\tfrac{ (1-y) (1-z)x}{(1-x) (1-x\,y\,z)}\right)
%id5.3t/
	+\text{permutations} =
%id5.3tr
3\zeta_3-3\zeta_2 \ln(x\,y\,z)
%id5.3tr/
	\,,
\end{multline}
where $\widetilde{L}_3(z)$ is defined in Eq. \eqref{eq:tildeL}.

Eq. \eqref{eq:goncharov1} is equivalent to the relation found by Goncharov \cite{goncharov1991classical}, but requires a rational variable change. In the notations of \cite[Eq. (16.97)]{lewin1991structural}\footnote{Mind the sign typo therein: $L_3(-b_i/a_{i-1})$ should be read as $L_3(b_i/a_{i-1})$.}, this change reads
\begin{equation}
	a_1=-\frac{(1 - y) x}{1 - x},\qquad
	a_2=-\frac{(1 - x) z}{1 - z},\qquad
	a_3=-\frac{(1 - z) y}{1 - y}\,.
\end{equation}
Note that Eqs. \eqref{eq:goncharov1} and \eqref{eq:goncharov2} are explicitly symmetric with respect to all permutations of $\{x,y,z\}$.
\section{Conclusion}

In the present paper we have introduced an algorithm of finding all possible arguments of $\Li_n$ functions with a prescribed branching locus. We provide several examples of using this algorithm for discovering the functional identities between these functions.\footnote{Identities in computer-readable form can be found in the ancillary file \textit{identities.m}.}
%
%Note that exactly the same method is applicable to a wider class of functions: harmonic polylogarithms with positive indices \cite{Remiddi:1999ew}. Finally, we remark that in the context of application to Feynman integrals, two following generalizations of the presented approach are very much desirable. First,

\acknowledgments I appreciate warm hospitality of University of Science and Technology of China, Hefei, where a part of this work was done. I am grateful to Yang Zhang and Andrei Pomeransky for the interest to the work and fruitful discussions. I am especially thankful to Andrei Pomeransky for emphasizing  the relation of the presented algorithm with the Stothers-Mason theorem and its generalizations. This work has been supported by Russian Science Foundation under grant 20-12-00205.

\bibliographystyle{JHEP}
\bibliography{Arguments.bib}
\end{document}